\documentclass[runningheads]{llncs}
\usepackage{amsmath}
\usepackage[T1]{fontenc}
\usepackage{graphicx}
\usepackage{url}

\usepackage[ruled,vlined]{algorithm2e}

\begin{document}
\title{Dynamic Graph Embedding Through Hub-aware Random Walks}
\author{Aleksandar Tom\v{c}i\'{c}\orcidID{0000-0002-3206-7600} \and
Milo\v{s} Savi\'{c}\orcidID{0000-0003-1267-5411} \and
Du\v{s}an Simi\'{c}\orcidID{0009-0003-9889-8313} 
\and Milo\v{s} Radovanovi\'{c}\orcidID{0000-0003-2225-7803}}

\authorrunning{A. Tom\v{c}i\'{c} et al.}

\institute{Department of Mathematics and Informatics, Faculty of Sciences,\\ University of Novi Sad\\
	Trg Dositeja Obradovi\'{c}a 4, 21000 Novi Sad, Serbia\\
	\email{\{atomic, svc, dusan.simic, radacha\}@dmi.uns.ac.rs}}

\maketitle              % typeset the header of the contribution
\begin{abstract}
The role of high-degree nodes, or hubs, in shaping graph dynamics and structure is well-recognized in network science, yet their influence remains underexplored in the context of dynamic graph embedding. Recent advances in representation learning for graphs have shown that random walk-based methods can capture both structural and temporal patterns, but often overlook the impact of hubs on walk trajectories and embedding stability. In this paper, we introduce DeepHub, a method for dynamic graph embedding that explicitly integrates hub sensitivity into random walk sampling strategies. Focusing on dynnode2vec as a representative dynamic embedding method, we systematically analyze the effect of hub-biased walks across nine real-world temporal networks. Our findings reveal that standard random walks tend to overrepresent hub nodes, leading to embeddings that underfit the evolving local context of less-connected nodes. By contrast, hub-aware walks can balance exploration, resulting in embeddings that better preserve temporal neighborhood structure and improve downstream task performance. These results suggest that hub-awareness is an important yet overlooked factor in dynamic graph embedding, and our work provides a foundation for more robust, structure-sensitive representation learning in evolving networks.

\keywords{Dynamic graphs  \and Graph embeddings \and Dynnode2vec \and Hubs.}
\end{abstract}
\section{Introduction}

Many real-world complex system studied across various scientific and engineering domains can be represented as graphs or networks, where nodes denote entities and edges represent their relationships or interactions. These systems are often dynamic, meaning their underlying network structure evolves over time through the addition or removal of nodes and edges~\cite{holme2012temporal}. Networks that capture such temporal changes are commonly referred to as \textit{dynamic} or \textit{temporal networks}.

Graph-structured data collected within complex dynamical systems can be leveraged to build predictive models using machine learning (ML) techniques~\cite{barros2021survey,khoshraftar2024survey}. Common problems include node classification, clustering, link prediction, attribute inference, anomaly detection and diffusion modeling. Broadly, there are three main strategies to develop such predictive models: (1) using graph-native ML algorithms like iterative classification or community detection, (2) applying traditional ML techniques designed for tabular data to graph embeddings, and (3) employing graph neural networks (GNNs).

This paper focuses on dynamic graph embedding algorithms. Unlike GNNs, which excel in task specific modeling of feature rich attributed graphs, graph embedding methods offer a task agnostic solution for learning general purpose graph representations, especially suitable for more common feature-sparse or non-attributed networks. Among these, random walk based dynamic embedding methods stand out due to their scalability, robustness and widespread adoption~\cite{barros2021survey}. The core idea behind these approaches is to perform random walks from each node, treating the resulting sequences of node identifiers as sentences. This formulation effectively transforms the graph embedding problem into an analogous text embedding task. Besides traditional machine learning problems, graph embeddings can also be utilized for similarity search on graphs and learning graph similarity functions~\cite{Nikolentzos2017}. 

In previous work dealing with static (non-evolving) graphs~\cite{Tomcic2024}, it was shown that hub-aware random walks based on the notion of ``good'' and ``bad'' hubs~\cite{radovanovic2010hubs} improve the intrinsic quality of embeddings used for node classification. In~\cite{SAVIC2023}, it has been shown that static graph embeddings can be improved by adjusting random walk hyper-parameters and transition biases according to NC-LID, which is a local intrinsic dimensionality (LID) measure for nodes in a graph~\cite{2021_NCLID_SISAP}. To the best of our knowledge, existing dynamic graph embedding methods based on random walks do not take into account structural properties of nodes, such as hubness and LID, when sampling random walks. In this paper, we propose DeepHub -- a novel method for creating dynamic graph embeddings. DeepHub relies on biased random walks with transition probabilities computed according to hubness measures (degree centrality to which various scaling mechanisms could be applied). 

The remainder of this paper is structured as follows. In Section~\ref{sec:related-work} we provide an overview of existing methods for dynamic graph embedding, with a particular focus on random walk based techniques. Section~\ref{sec:hub-aware} presents DeepHub, detailing the methodology for incorporating hub awareness into random walk strategies to enhance the quality of node representations. In Section~\ref{sec:eval-metho} we describe our experimental setup, including datasets, evaluation metrics and the baseline method (Dynnode2vec~\cite{mahdavi2018dynnode2vec}) used for comparison. The analysis of obtained experimental results is given in Section~\ref{sec:expe-res}. Finally, Section~\ref{sec:conc} summarizes the key findings of the paper and outlines potential directions for future research. 

\section{Related Work}
\label{sec:related-work}

Dynamic graph embedding methods aim to learn low-dimensional representations of nodes in evolving networks, while preserving structural and temporal information. Among these, a prominent class of techniques is based on random walks performed on individual graph snapshots. These methods typically generate node contexts via random walks and subsequently learn embeddings using models such as node2vec~\cite{grover2016node2vec}. 

A key limitation of this approach lies in the handling of temporal dependencies. Specifically, while the random walks are executed independently on each graph snapshot, the temporal linkage between the embedding matrices of successive snapshots is not captured during the walk process itself. Instead, temporal consistency is introduced afterward through various mechanisms that enforce smooth transitions in the learned representations across time. For instance, embeddings can be computed separately for each snapshot using static methods like node2vec or DeepWalk~\cite{deepwalk}. Then, temporal relationships between these embeddings are enforced post hoc via techniques ranging from simple vector concatenation to more sophisticated alignment strategies. The methods proposed in~\cite{de2018combining} and \cite{mitrovic2019dyn2vec} apply vector concatenation to link node representations across time, with the former using node2vec and the latter relying on a DeepWalk variant that biases edge selection based on normalized edge weights. Other approaches, such as those presented in~\cite{chen2019dynamic}, leverage probabilistic priors (e.g., Gaussian) and model the evolution of embeddings using dynamic Bernoulli distributions. Similarly, tNodeEmbed~\cite{singer2019node} preserves local static neighborhoods using node2vec, then aligns node embeddings across time using the Orthogonal Procrustes method, and employs an LSTM to optimize task-specific objectives such as link prediction and node classification. DynSEM~\cite{zhou2019dynamic} also performs static node2vec embedding on each snapshot, followed by alignment using Orthogonal Procrustes, and finally optimizes a joint loss function that includes temporal smoothness constraints. These strategies reflect a common pattern in the literature: temporal coherence is typically not embedded within the random walk phase but enforced afterward through alignment or regularization mechanisms.

Generating random walks at every timestamp is computationally expensive and often impractical for large-scale or high-frequency dynamic networks. To mitigate this, several approaches initialize node embeddings at the first snapshot using a static random walk-based method and subsequently update them incrementally. These methods leverage the insight that only a small subset of nodes typically undergoes significant structural changes over time. The most representative of them, Dynnode2vec~\cite{mahdavi2018dynnode2vec}, realizes this strategy by sampling node sequences solely for the evolving nodes at each timestamp.

In contrast to approaches that update embeddings incrementally, other methods explicitly incorporate temporal dependencies into the random walk process itself. In particular, CTDNE~\cite{CTDNE2018} is a method for learning time-preserving node representation by constraining random walks to respect temporal order. Rather than treating the dynamic graph as a sequence of static snapshots, CTDNE performs temporal random walk on an aggregated graph constructed from the edge stream, ensuring that each transition from node $u$ to a node $v$ only occurs if the edge to $v$ has a timestamp greater than or equal to that of the previous edge.

Another method, STWalk~\cite{STWALK2018}, approaches the temporal representation problem by modeling node trajectories across discrete graph snapshots. Given a sequence of graph snapshots, STWalk generates embeddings by performing two types of walks: one over the current spatial neighborhood of a node and another over its own historical states. Specifically, the embedding of node $u$ at time $t$ is optimized to preserve proximity to both its neighbors at $t$ and its own embeddings from earlier timestamps.  

\section{Hub-aware Random Walk Methods}
\label{sec:hub-aware}

In this section, we present our approach, describing how hub-awareness is incorporated in random walk strategies to improve the quality of node representation.

Let $G$ denote a dynamic graph that is represented by an evolutionary sequence of graph snapshots denoted by $G_{i}$ ($1 \leq i \leq N$, where $N$ is the total number of snapshots or graph evolutionary steps). Following the standard notation, nodes and links in $G_{i}$ are denoted by $V_{i}$ and $E_{i}$, respectively.  

In general, dynamic graph embedding algorithms perform two steps:
\begin{enumerate}
\item train an initial text embedding model from random walks sampled on $G_{1}$,
\item update the model on random walks sampled from each subsequent graph snapshot $G_{i}$ ($2 \leq i \leq N$). 
\end{enumerate}
This generic process is shown in Algorithm 1. The embedding of $G$, in the pseudo-code denoted by $\hat{E}$, is a sequence (array or list) of singleton embeddings, one singleton embedding per graph snapshot. Two important constituents of the algorithm are (1) the sampling strategy, denoted by $S$, and (2) the function determining so-called delta (or evolving) nodes, denoted by $D$. $S$ is applied to $G_{i}$ to obtain random walks from starting nodes that are determined by $D$ when updating the model (denoted by $M$). $D$ considers two consecutive graph snapshots to determine nodes passing through some evolutionary changes. In Dynnode2vec~\cite{mahdavi2018dynnode2vec} the set of delta nodes is defined as the union of ``new'' nodes (nodes appearing in $V_{i}$ that are not present in $V_{i - 1}$) and nodes whose ego networks have changed in the transition from $G_{i - 1}$ to $G_{i}$. An ego network refers to a node and its immediate neighbors along with the edges connecting them. After each model update, the singleton embedding of $G_{i}$ is extracted from $M$ and added to $\hat{E}$. 

\begin{algorithm}[htb!]
	\label{alg1}
	\small
	\SetAlgoLined
	\DontPrintSemicolon
	\SetKwInOut{Input}{input}
	\SetKwInOut{Output}{output}
	\Input{$G = <G_{1}, G_{2} ..., G_{N}>$ -- input dynamic graph\\
		$S$ -- sampling strategy\\
		$D$ -- delta nodes function\\
	}
	\Output{$\hat{E}$ -- an embedding of $G$} 
	
	\BlankLine \BlankLine 
	
	$T$ = sample-random-walks($G_{1}, V_{1}, S$) \\
	$M$ = train-model($T$) \\
	$\hat{E} = [\text{extract-embedding}(M)]$
	\BlankLine
	
	\For{$i = 2$ {\bf to} $N$} {
		$d$ = $D$($G_{i}$, $G_{i - 1}$) \\
		$T$ = sample-random-walks($G_{i}, d, S$) \\
		$M$ = update-model($M, T$) \\
		$\hat{E}$.append(extract-embedding($M$)) \\
	}
	
	\BlankLine
	{\bf return} $\hat{E}$
	
	\caption{{\bf General form of random walk dynamic graph embedding algorithms}}	
	\SetAlgoRefName{alg1}
	\SetAlgoCaptionSeparator{'.'}
\end{algorithm}

Random walk sampling can be also described in a generic way as shown in Algorithm 2. Besides the input graph, the set of delta nodes corresponding to starting nodes and the sampling strategy, the algorithm has two additional parameters that are functions. For node $v$, $F_{nw}(v)$ returns the number of random walks sampled from $v$, while $F_{wl}(v)$ returns the length of random walks sampled from $v$. It is important to emphasize that $F_{nw}$ and $F_{wl}$ in the existing state-of-the-art methods are actually constants. 

\begin{algorithm}[htb!]
	\label{alg2}
	\small
	\SetAlgoLined
	\DontPrintSemicolon
	\SetKwInOut{Input}{input}
	\SetKwInOut{Output}{output}
	\Input{$G$ -- input graph\\
		$W$ -- delta nodes ($W \subseteq V$) \\
		$F_{nw}$ -- function determining the number of walks per starting node \\
		$F_{wl}$ -- function determining walk length \\
		$S$ -- sampling strategy\\
	}
	\Output{$T$ -- list of random walks} 
	
	\BlankLine \BlankLine 
	
	$T$ = [] \\
	\For{$v \in W$} {
		\For{$i = 1$ {\bf to} $F_{nw}(v)$} {
			$s$ = [] \\
			$c$ = $v$ \\
			\For{$j = 1$ {\bf to} $F_{wl}(v)$} {
				$s$.append({\it id} of $c$) \\
				$NC$ = the set of neighbors of $c$ \\ 
				$next$ = select a node from NC according to $S$\\ 
				$c = next$ \\
			}
			$T$.append($s$)
		}
	}
	
	\BlankLine
	{\bf return} $T$
	\caption{{\bf General form of random walk sampling}}	
	\SetAlgoRefName{alg2}
	\SetAlgoCaptionSeparator{'.'}
\end{algorithm}

While traditional random walk methods for temporal graphs, such as Dynnode2vec, rely on uniform or simple biased strategies for neighbor selection, the DeepHub algorithm introduces a hub-aware random walk sampling mechanism. The key motivation behind DeepHub is to better capture the structural significance of hubs. 

DeepHub extends the random walk paradigm by incorporating degree centrality. At each step of a random walk, with a conditional probability, neighbors of the current node are scored based on their degree, with two flexible configuration modes. The default hub-biased mode favors transitions towards high degree neighbors. In this mode the probability that a neighbor will be visited is proportional to its degree centrality. Alternatively, the inverse mode prioritizes low degree nodes (probabilities are inversely proportional to degree centrality). An optional logarithmic scaling can be applied to smooth degree differences, mitigating the dominance of very high degree nodes.

In addition to degree based selection, DeepHub integrates two parameters. The first parameter, backtracking probability $p$, allows the walker to return back to the previous node in the walk. The second parameter, random move probability $u$, enables uniform transition to one of the neighbors, i.e., each of the neighbors has the same probability to be selected as the next node in the walk. These mechanisms introduce variability in walk trajectories, supporting the discovery of both persistent and transient temporal structures.

\begin{algorithm}[htb!]
	\caption{\textbf{DeepHub} strategy: selecting next node in a walk using degree-based scoring with optional inverse and logarithmic scaling.}
	\label{algUnifications}
	\small
	\SetAlgoLined
	\DontPrintSemicolon
	\SetKwInOut{Input}{input}
	\SetKwInOut{Output}{output}
	
	\Input{
		$start\_node$, $current\_node$, $prev\_node$ -- walk context \\
		$NC$ -- neighbors of $current\_node$ \\
		$p$, $u$ -- probabilities for backtracking and random move \\
		$G$ -- input graph \\
		$inverse$, $log\_scaling$ -- boolean configuration flags
	}
	\Output{$n$ -- next node}
	
	\BlankLine
	Generate random number $r \in [0, 1]$ \\
	\If{$r < p$}{
		\Return $prev\_node$
	}
	
	Generate random number $r \in [0, 1]$ \\
	\If{$r < u$}{
		\Return random node from $NC$ uniformly selected
	}
	
	\BlankLine
	Initialize $P \gets []$ \\
	$max\_deg \gets \max\limits_{n \in NC} \deg(n)$ \\
	
	\ForEach{$n \in NC$}{
		\uIf{$inverse$}{
			$score \gets 1 + max\_deg - \deg(n)$
		}
		\Else{
			$score \gets 1 + \deg(n)$
		}
		\If{$log\_scaling$}{
			$score \gets \log(1 + score)$
		}
		Append $score$ to $P$
	}
	
	Normalize $P$ to form a probability distribution \\
	Select $n$ from $NC$ using $P$ as probabilities \\
	\Return $n$
	
\end{algorithm}

For a given walk context, DeepHub first probabilistically determines whether to backtrack (to behave as node2vec). If backtracking is not selected, then it tries to perform an uniform transition (to behave as DeepWalk). If neither occurs, it computes degree based scores for the current node's neighbors, adjusted according to the chosen mode and scaling. These scores are normalized into a probability distribution, from which the next node is sampled. We also use Laplace smoothing (adding 1 to degree centrality) in order to avoid zero probabilities in case of the logarithmic scaling.

\section{Evaluation Methodology}
\label{sec:eval-metho}

The primary goal of general-purpose graph embedding methods is to preserve the structural properties of networks in the embedded space. Effective graph embedding techniques should produce embeddings that allow for the reconstruction of the original graphs. The graph reconstruction process involves computing the Euclidean distance between the embedding vectors of each pair of nodes and then connecting the closest $|E|$ node pairs, where $|E|$ represents the number of edges in the original graph. Let $n$ denote an arbitrary node and $C$ be the number of correctly reconstructed links incident to $n$. The following embedding evaluation metrics are used to assess the quality of the embeddings:
\begin{itemize}
	\item Precision -- $C$ divided by the number of links $n$ has in the reconstructed graph.
	\item Recall -- $C$ divided by the number of links $n$ has in the original graph.
	\item $F_1$ score -- The harmonic mean of precision and recall.
\end{itemize}
Higher values of precision, recall, and $F_1$ indicate fewer link reconstruction errors for node $n$. At the graph level, precision, recall, and $F_1$ scores can be obtained by micro-averaging over all nodes. 

The baseline method, Dynnode2vec, is tuned by selecting the values of its hyperparameters $p$ (the return back parameter) and $q$ (the in-out parameter) that maximize the $F_1$ score in the last snapshot. We consider five different graph embedding dimensions (10, 25, 50, 100, and 200), and values for $p$ and $q$ from the set \{0.25, 0.50, 1, 2, 4\}. The number of random walks per node and the length of each walk are fixed to 10 and 32, respectively. For each configuration of hyperparameters, the $F_1$ score is determined by averaging the results of 10 independent runs.

To investigate the relationship between node degree and the quality of dynamic graph embeddings, we employ a methodology similar to those proposed in~\cite{Knezevic2025} (the methodology from~\cite{Knezevic2025} examines relationships between NC-LID and $F_{1}$ scores in the dynamic graph setting). More concretely, as the first step, we compute Spearman correlations between node degrees and the embedding evaluation metrics defined above, across all graph snapshots. Strong negative correlations between node degrees and embedding metrics suggest that degree centrality can highlight nodes with poorly constructed embeddings.

Going forward in the evaluation methodology, we need to introduce the following definition.

\begin{definition} 
	\label{definition-hub}	
	\upshape[\textbf{Hub}] 
	Let $V$ be a set od nodes in a graph (or more precisely, in one snapshot of a dynamic graph) and let $k_{v}$ denote the degree of node $v \in V$. Node $h$ is a hub if it belongs to the set of hubs $H$. $H$ is the \textbf{minimal} set satisfying the following condition
	$$\sum\limits_{h \in H} k_{h} > \sum\limits_{o \in V \setminus H} k_{o}.$$ 
\end{definition}

In other words, $H$ is the minimal set of high-degree nodes whose total degree is 
higher than the total degree of the rest of the nodes in the graph.

Furthermore, based on Definition~\ref{definition-hub}~\cite{savic2017analysis}, for each snapshot, nodes are categorized into two groups: \textit{hubs} and \textit{non-hubs}. Using this partitioning, we compare the $F_1$ scores of hub and non-hub nodes to investigate the impact of node centrality on embedding intrinsic quality. The comparison is carried out using the Mann-Whitney U (MWU) test~\cite{mann1947test}. The MWU test checks the null hypothesis that the $F_1$ scores for group \textit{hubs} do not tend to be either greater or smaller than those in group \textit{non-hubs}. Additionally, we calculate two probabilities of superiority (PS), which measure the strength of stochastic inequality:
\begin{itemize}
	\item PS(hubs): The probability that a randomly selected \textit{hub} has a higher $F_1$ score than a randomly selected \textit{non-hub}.
	\item PS(non-hubs): The probability that a randomly selected \textit{non-hub} has a higher $F_1$ score than a randomly selected \textit{hub}.
\end{itemize}

If the null hypothesis of the MWU test is rejected ($p$-value below 0.05), we can state that there are statistically significant differences between \textit{hubs} and \textit{non-hubs}. If $PS_{hubs} \gg PS_{non-hubs}$, hubs tend to have intrinsically better node embedding vectors, while for $PS_{non-hubs} \gg PS_{hubs}$ the exact opposite holds.

\section{Experiments and Results}
\label{sec:expe-res}

Our experimental evaluation of DeepHub is conducted on real-world dynamic networks listed in Table~\ref{datasets}. Selected networks are widely used in the literature related to dynamic graphs and they can be retrieved from two publicly-available repositories: SNAP\footnote{\url{https://snap.stanford.edu/data/#temporal}} and Network Repository.\footnote{\url{https://networkrepository.com/dynamic.php}} Table~\ref{datasets} shows the total number of nodes ($|V|$) and the total number of edges ($|E|$) that are present throughout network evolution, the number of snapshots and time resolution of each snapshot. It also gives information about the average activation of nodes and edges, denoted by $a(V)$ and $a(E)$, respectively. The activation of a node (resp. edge) is the number of snapshots in which the node (resp. edge) is present or active.

\begin{table}[t!]
	\centering
	\setlength{\tabcolsep}{6pt}
	\caption{Experimental datasets.}
	\label{datasets}
	\begin{tabular}{lllllll}
		\hline
		Network & $|V|$ & $|E|$ & snapshots & time resolution & $a(V)$ & $a(E)$ \\ \hline
		ia-hospital & 75 & 1369 & 4 & day & 2.79 & 1.34 \\ 
		ia-contacts & 113 & 2470 & 3 & day & 2.65 & 1.19 \\ 
		ia-enron & 151 & 2047 & 11 & 3-months & 6.57 & 1.74 \\ 
		radoslaw-email & 167 & 4519 & 9 & month & 7.5 & 2.35 \\ 
		ia-primschool & 242 & 9843 & 2 & day & 1.98 & 1.16 \\ 
		fb-forum & 898 & 8212 & 5 & month & 3.03 & 1.28 \\ 
		email-eu & 986 & 22205 & 17 & month & 12.29 & 2.79 \\ 
		college-msg & 1894 & 14474 & 6 & month & 2.23 & 1.08 \\ 
		fb-messages & 1899 & 16415 & 7 & month & 2.45 & 1.07 \\ \hline
	\end{tabular}
\end{table}

The results of Dynnode2vec (DN2V) tuning are presented in Table~\ref{dn2v}. This table shows the maximal $F_1$ scores for dynamic networks from our experimental corpus, the dimension in which the maximal $F_{1}$ score is achieved (Dim.), precision, recall and the corresponding values of $p$ and $q$. It can be observed that for all dynamic networks we have $F_1$ scores in the range [0.29, 0.78]. Such high $F_1$ scores imply that Dynnode2vec embeddings preserve the structure of input dynamic networks to a great extent. Additionally, it can be seen that precision and recall tend to have close values implying that the corresponding quality attributes tend to be balanced in the best Dynnode2vec embeddings. Consequently, it can be concluded that Dynnode2vec achieves very good performance at making general-purpose vector representations of nodes in dynamic networks, justifying our choice of this particular state-of-the-art dynamic graph embedding method for the purpose of DeepHub evaluation. We also empirically evaluated other dynamic graph embedding methods (CTDNE~\cite{CTDNE2018} and STWalk~\cite{STWALK2018}), but on our experimental datasets they exhibit noticeably higher graph reconstruction errors (lower $F_{1}$ scores with hyper-parameter tuning included).

\begin{table}[b!]
	\centering
	\setlength{\tabcolsep}{6pt}
	\caption{Characteristics of the best Dynnode2vec embeddings.}
	\begin{tabular}{lllllll}
		\hline
		Network & Dim. & p & q & Precision & Recall & $F_1$ \\ \hline
		radoslaw-email & 25 & 0.5 & 2 & 0.6373 & 0.5973 & 0.6165 \\
		college-msg & 50 & 4 & 1 & 0.5238 & 0.5669 & 0.5438 \\ 
		email-eu & 50 & 0.5 & 0.5 & 0.5623 & 0.6076 & 0.5839\\
		fb-forum & 200 & 1 & 0.5 & 0.5845 & 0.5943 & 0.5889 \\
		fb-messages & 100 & 2 & 2 & 0.2805 & 0.3121 & 0.2922\\ 
		ia-contacts & 200 & 0.5 & 4 & 0.6309 & 0.5314 & 0.5766 \\ 
		ia-enron & 25 & 0.5 & 2 & 0.3774 & 0.4252 & 0.3985 \\ 
		ia-hospital & 25 & 0.5 & 4 & 0.6183 & 0.5799 & 0.5981 \\ 
		ia-primary & 200 & 0.5 & 4 & 0.7919 & 0.7730 & 0.7823 \\ \hline
	\end{tabular}
	\label{dn2v}
\end{table}

\begin{figure}[tb]	
	\centering
	\includegraphics[width=0.98\textwidth]{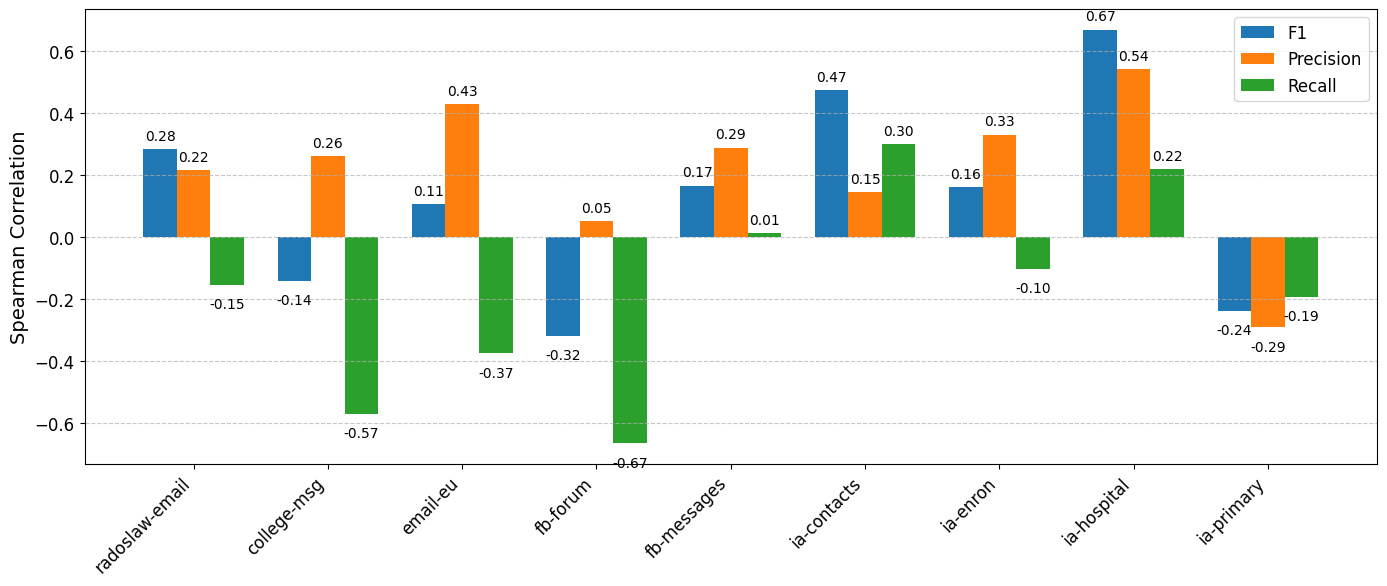}
	\caption{Spearman correlations between node's degree and graph embedding evaluation measures ($F_1$, precision and recall) for DN2V.}
	\label{fig-dn2v}
\end{figure}

To assess the relationship between local structural properties and embedding quality, we computed Spearman correlation coefficients between node's degree and the embedding evaluation metrics ($F_1$, Precision, and Recall) across all datasets using the best Dynnode2vec embeddings. As illustrated in Figure~\ref{fig-dn2v}, the strength and direction of correlations vary across datasets. Notably, positive correlations with $F_1$ appear in \textit{ia-hospital} (0.6693), \textit{ia-contacts} (0.4744), and \textit{radoslaw-email} (0.2837). In contrast, datasets such as \textit{fb-forum} and \textit{college-msg} show strong negative correlations. The presence of such moderate to strong positive or negative correlations imply that there are notable differences in the embedding quality of high degree nodes (hubs) and low degree nodes (non-hubs). Positive correlations mean that hubs tend to have better embedding vectors, whereas negative correlations imply the opposite.   

Table~\ref{tab:dn2v} summarizes the results of the Mann-Whitney U (MWU) statistical tests along with the associated probabilities of superiority. $F_1$(hubs) and $F_1$(non-hubs) represent the average $F_1$ score for hub nodes and non-hub nodes, respectively. $U$ is the value of the MWU test statistic, and the null hypothesis of the test is accepted if the $p$-value of $U$ is higher than 0.05, as indicated in the ACC column of Table~\ref{tab:dn2v}. A ``no'' in the ACC column means that the null hypothesis is rejected, suggesting statistically significant differences between hub and non-hub nodes with respect to their $F_1$ scores. The results reveal that there are statistically significant differences in $F_1$ scores between hub and non-hub nodes in 8 out of the 9 networks (the null hypothesis is accepted only for \textit{ia-primary}). In 4 networks we have that $F_1$(hubs) < $F_1$(non-hubs) and PS(hubs) $\ll$ PS(non-hubs), indicating that the $F_1$ scores of hub nodes tend to be significantly lower than those of non-hub nodes. Thus, it can be concluded that hubness of nodes in some cases have a positive impact and in some cases a negative impact on intrinsic embedding quality.

\begin{table}[htb!]
	\centering
	\caption{Comparison of $F_1$ scores of hub nodes and non-hub nodes using the Mann-Whitney U test for DN2V.}
	\begin{tabular}{llllllllllllll}
		\hline
		Network & $F_1$(hubs) & $F_1$(non-hubs) & $U$ & $p$ & ACC & PS(hubs) & PS(non-hubs) \\
		\hline
		radoslaw-email & 0.6436 & 0.4948 & 203919.5 & 0.0 & no & 0.6907 & 0.306 \\
		college-msg & 0.2878 & 0.4024 & 848726 & 0.0 & no & 0.399 & 0.5705 \\
		email-eu & 0.4821 & 0.4965 & 10632201 & 0.0019 & no & 0.4747 & 0.5165 \\
		fb-forum & 0.3723 & 0.4917 & 463495 & 0.0 & no & 0.3846 & 0.5878 \\
		fb-messages & 0.2804 & 0.2484 & 1620864.5 & 0.0 & no & 0.485 & 0.3686 \\
		ia-contacts & 0.6405 & 0.4667 & 14197.5 & 0.0 & no & 0.7488 & 0.2486 \\
		ia-enron & 0.364 & 0.296 & 108354 & 0.0 & no & 0.5531 & 0.3818 \\
		ia-hospital & 0.7245 & 0.4715 & 7898 & 0.0 & no & 0.8497 & 0.1475 \\
		ia-primary & 0.7681 & 0.771 & 24495.5 & 0.3277 & yes & 0.4721 & 0.5265 \\
		\hline
	\end{tabular}
	
	\label{tab:dn2v}
\end{table}

Table~\ref{tab:deep_hub} repeats the same evaluation for DeepHub embeddings. Compared to DN2V, the results show that DeepHub reduces the differences in $F_1$ scores between hub and non-hub nodes in several networks, with the null hypothesis of equal distributions now accepted in 3 out of 9 networks (\textit{radoslaw-email}, \textit{email-eu}, and \textit{ia-contacts}). This indicates improved embedding fairness with respect to hubness in these cases.

\begin{table}[b!]
	\centering
	\caption{Comparison of $F_1$ scores of hub nodes and non-hub nodes using the Mann-Whitney U test for DeepHub.}
	\begin{tabular}{llllllllllllll}
		\hline
		Network & $F_1$(hubs) & $F_1$(non-hubs) & $U$ & $p$ & ACC & PS(hubs) & PS(non-hubs) \\
		\hline
		radoslaw-email & 0.699 & 0.6634 & 155374.5 & 0.14 & yes & 0.5253 & 0.4702 \\
		college-msg & 0.3125 & 0.42 & 838605 & 0.0 & no & 0.3949 & 0.5763 \\
		email-eu & 0.5088 & 0.5093 & 11040627 & 0.7128 & yes & 0.4931 & 0.498 \\
		fb-forum & 0.3907 & 0.4968 & 471752 & 0.0 & no & 0.3963 & 0.5853 \\
		fb-messages & 0.2741 & 0.2569 & 1566147.5 & 0.0003 & no & 0.463 & 0.3843 \\
		ia-contacts & 0.6276 & 0.581 & 10622.5 & 0.0922 & yes & 0.5603 & 0.4379 \\
		ia-enron & 0.4132 & 0.3531 & 105461.5 & 0.0008 & no & 0.5487 & 0.4086 \\
		ia-hospital & 0.6589 & 0.5782 & 6331 & 0.0 & no & 0.6805 & 0.3161 \\
		ia-primary & 0.776 & 0.8011 & 18648.5 & 0.0 & no & 0.3594 & 0.6396 \\
		\hline
	\end{tabular}
	
	\label{tab:deep_hub}
\end{table}

\begin{figure}[htb!]
	\centering
	\includegraphics[width=0.98\textwidth]{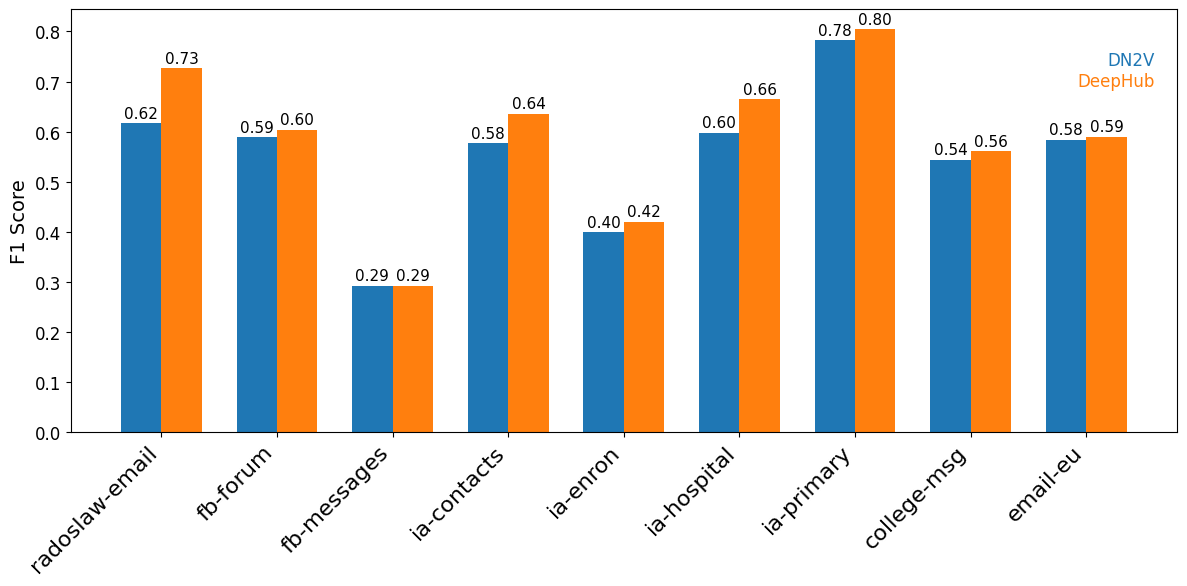}
	\caption{Comparison of $F_{1}$ scores between DN2V and DeepHub.}
	\label{fig:deephub-bars1}
\end{figure}

The comparison of $F_{1}$ scores between DeepHub and Dynnode2vec for all experimental datasets is shown in Figure~\ref{fig:deephub-bars1}. The figure also gives numerical values of $F_{1}$ for both DeepHub and the baseline method. It can be seen that DeepHub achieves higher $F_{1}$ scores for 8 out of 9 networks (all networks except \textit{fb-messages}). Considerable improvements in $F_{1}$ are present for \textit{radoslaw-email} ($F_{1}$ increased from 0.62 to 0.73), \textit{ia-contacts} ($F_{1}$ increased from 0.58 to 0.64), and \textit{ia-hospital} ($F_{1}$ increased from 0.6 to 0.66). This result implies that (1) DeepHub provides dynamic graph embeddings having a good degree of intrinsic quality ($F_{1}$ scores higher than 0.5 in most of the cases), and (2) DeepHub outperforms the baseline method.

A more detailed comparison of DeepHub and Dynnode2vec is given in Table~\ref{tab:deephub}. Besides $F_{1}$ scores, this table provides information about the best values of DeepHub parameters $p$ (backtrack probability) and $u$ (uniform sampling probability) and the type of scaling applied to degree centrality (it can be normal, log, inverse or inverse-log; normal means no transformation). DeepHub outperforms Dynnode2vec in nearly all datasets, with the exception of \textit{fb-messages}. In this case, the $F_1$ score experiences a marginal decrease of 0.0003, corresponding to -0.09\% relative drop, which does not reflect a significant performance decline.

\begin{table}[b!]
	\centering
	\setlength{\tabcolsep}{4pt}
	\caption{Best $F_1$-scores for DeepHub algorithm and its comparison with DN2V.}
	\begin{tabular}{lllllllr}
		\hline
		Network & DN2V & DeepHub & $p$ & $u$ & type & $F_1$-diff & $F_1$-imp [\%]\\ \hline
		radoslaw-email & 0.6165 & 0.7261 & 0.5 & 0.5 & inverse & 0.1096 & 17.7760 \\ 
		fb-forum & 0.5889 & 0.6040 & 0.15 & 0.15 & inverse-log & 0.0151 & 2.5586 \\ 
		fb-messages & 0.2923 & 0.2920 & 0.15 & 0.15 & inverse-log & -0.0003 & -0.0926 \\ 
		ia-contacts & 0.5766 & 0.6355 & 0.5 & 0.5 & inverse & 0.0589 & 10.2203 \\ 
		ia-enron & 0.3986 & 0.4196 & 0.5 & 0.5 & inverse-log & 0.0210 & 5.2626 \\ 
		ia-hospital & 0.5982 & 0.6644 & 0.5 & 0.5 & inverse & 0.0662 & 11.0689 \\ 
		ia-primary & 0.7823 & 0.8047 & 0.25 & 0.25 & inverse-log & 0.0224 & 2.8606 \\ 
		college-msg & 0.5439 & 0.5607 & 0 & 0.15 & inverse-log & 0.0168 & 3.0970 \\ 
		email-eu & 0.5839 & 0.5901 & 0.15 & 0.15 & inverse-log & 0.0062 & 1.0654 \\ \hline
	\end{tabular}
	\label{tab:deephub}
\end{table}

For all networks except \textit{college-msg}, the optimal value of $p$ is higher than~0. This implies that returning back to the previous node in the random walk to a certain degree is an important factor when sampling biased random walks. The same holds for $u$ ($u > 0$ for all networks), i.e., biased random walks should include uniform selection of neighboring nodes to a certain extent. The type of transformation applied to degree centrality is either inverse-log (6 networks) or inverse (3 networks). This means that in all networks biased random walks tend to avoid hubs.  

The obtained experimental findings demonstrate that DeepHub is a robust and adaptable temporal graph embedding method that consistently enhances intrinsic embedding quality across a wide range of dynamic network datasets. The method proves effective not only for small social networks but also for large, temporally rich graphs, showcasing its scalability and ability to adjust to different structural and temporal characteristics. The consistent performance improvement across most datasets, irrespective of domain or connectivity patterns, underscores the versatility of DeepHub. Moreover, the minimal performance drop observed in only one dataset highlights its stability, making it a potentially reliable approach for machine learning tasks such as link prediction or network classification in evolving environments.

\section{Conclusions and Future Work}
\label{sec:conc}
\vspace{-5pt}

This paper introduced DeepHub, a dynamic graph embedding method that enhances random walk based node representations through a hub-aware selection mechanism. By adjusting transition probabilities using inverse or logarithmic-inverse degree-based scoring, DeepHub mitigates hub dominance and captures more informative structural patterns in feature sparse and evolving networks. The experimental results presented in this paper show that DeepHub outperforms Dynnode2vec. Thus, the method provides a flexible, task agnostic approach for generating robust node embeddings of dynamic graphs, additionally offering a scalable solution for variety of graph-based machine learning tasks due to its lightweight nature.  

For future work, we aim to see how DeepHub performs in concrete applicative tasks such as link prediction and diffusion modeling in dynamic networks. One of the goals will be to examine the applicability of DeepHub embeddings to analyze dynamical processes on networks in scenarios like node influence propagation, misinformation detection and epidemic spreading.

Besides methods based on random walks, another large and important category of dynamic graph embedding methods are those based on autoencoders~\cite{barros2021survey}. In this study we have shown that hubness plays an important role when forming dynamic graph embeddings. Another line of our future work will be to design and evaluate hub-aware autoencoder methods, following the idea of incorporating hubness metrics into the autoencoder loss function.  
\vspace{-5pt}

\begin{credits}
\subsubsection{\ackname}
This research is supported by the Science Fund of the Republic of Serbia, \#7462, Graphs in Space and Time: Graph Embeddings for Machine Learning in Complex Dynamical Systems -- TIGRA. The authors would like to thank the anonymous reviewers for their insightful suggestions and comments that helped improve the quality of the paper.
\vspace{-5pt}

\subsubsection{\discintname}
The authors have no competing interests to declare that are relevant to the content of this article.
\end{credits}
\vspace{-8pt}

\bibliographystyle{splncs04}

\bibliography{hubrefs}

\end{document}